\documentclass[a4paper,conference]{IEEEtran}

\usepackage[utf8]{inputenc}
\usepackage[english]{babel}

\def\mytitle{Smoothed Particle Magnetohydrodynamics: A State of the Union}
\def\myauthor{Benjamin T. Lewis, Matthew R. Bate, \& Terry S. Tricco}

\usepackage{cite}      

\usepackage{graphicx}  

\usepackage{subfigure} 

\usepackage{amsmath}   

\usepackage{amssymb} 

\usepackage{aas_macros}
\usepackage{textcomp}
\usepackage{gensymb}
\usepackage{empheq}
\usepackage{csquotes}
\usepackage[hidelinks, pdfauthor={\myauthor}, pdftitle={\mytitle}, pagebackref=false]{hyperref}
\usepackage{cleveref}

\usepackage{color,soul} 
\def\diff{} 

  \crefname{figure}{Fig.}{Figs.}
  \Crefname{figure}{Fig.}{Figs.}
  \crefname{section}{$\S$}{$\S\S$}
  \Crefname{section}{$\S$}{$\S\S$}
  \crefname{equation}{Eqn.}{Eqns.}
  \Crefname{equation}{Eqn.}{Eqns.}

\usepackage{fancyheadings}
\def\rmn{\mathrm}

\begin{document}

\title{Smoothed Particle Magnetohydrodynamics: \\ A State of the Union }

\author{\IEEEauthorblockN{Benjamin T. Lewis, Matthew R. Bate}
\IEEEauthorblockA{School of Physics \& Astronomy\\
University of Exeter\\
Exeter, EX4 4QL\\
United Kingdom\\
blewis@astro.ex.ac.uk}
\and
\IEEEauthorblockN{Terrence S. Tricco}
\IEEEauthorblockA{Canadian Institute for Theoretical Astrophysics\\
University of Toronto\\
60 St. George Street, 14th floor\\
Toronto, ON. M5S 3H8\\
Canada}
}


\maketitle

\begin{abstract}
Obtaining a stable magnetohydrodynamical (MHD) formalism in SPH -- i.e. smoothed particle magnetohydrodynamics (SPMHD) -- has proven remarkably difficult. To implement MHD requires two steps: a modification to the momentum equation and an induction equation, and both present challenges. We first provide an overview of how SPMHD is implemented, and then discuss how this implementation fails and the limitation of various corrective methods -- with particular reference to the effects of particle disorder.
\diff{Although there are many problems for which, with careful choice of corrective measures, good results can be obtained, we then show that, at the very limits of the state of the art, the ability to perform stable MHD calculations in SPH is curtailed by numerical issues.} 
\end{abstract}

\section{Introduction}
\label{sec:intro}
The invention of smoothed particle hydrodynamics (SPH) by \cite{1977AJ.....82.1013L} and \cite{1977MNRAS.181..375G} was almost immediately followed by attempts to extend the method from a purely hydrodynamic (HD) regime to include ideal magnetohydrodynamics (MHD). Unlike fluids where the force is provided primarily by the hydrodynamic pressure (usually via an equation of state from the density and internal energy), \diff{in MHD the force is also related to the magnetic field strength and velocity which makes smoothed particle magnetohydrodynamics (SPMHD) remarkably resistant to a robust implementation.}

SPH is ideal for many large astrophysical calculations, which regularly involve density ratios of many orders of magnitude, for example, in our protostellar collapse calculations a density ratio between the initial conditions and the inner region of the disc of over $10^{7}$ is normal. This presents a challenging problem for many Eulerian grid codes but with an adaptive smoothing length method is easily tractable in SPH. MHD is difficult in any numerical regime, and the particulate nature of SPH is especially troublesome, but the inherent HD advantages mean a stable and correct SPMHD method is extremely valuable for computational astrophysics -- astrophysical plasmas being almost invariably magnetised to some degree \cite{1949Sci...109..165H}.

Early work, e.g. \cite{1985MNRAS.216..883P}, considered this and discovered that in addition to the ubiquitous computational problem in maintaining a solenoidal magnetic field \diff{(\textit{i.e.} one in which }$\nabla^iB^i = 0$\diff{, where }$B^i$\diff{ is the magnetic field vector)}, SPMHD was also susceptible to an instability whereby SPH particles would unphysically clump together if the magnetic tension force was capable of overhauling the fluid pressure, similar to how external strains can cause issues -- manifesting as negative pressure terms -- in solid calculations \cite{1995JCoPh.116..123S}. 

Nonetheless, many challenging calculations have been performed with success provided the fluid stays within the abilities of the corrective measures used -- or withing a regime which can maintain stability and correctness without such corrections. These calculations range from our protostellar collapse simulations \cite{2009MNRAS.398...33P}, \cite{2014MNRAS.437...77B}, \cite{2015MNRAS.451.4807L}, and \cite{LewisSubmit} which model the formation of a protostar from an initial molecular cloud core, to merging neutron stars \cite{2007MNRAS.379..915R}, and simulations of the evolution of whole molecular clouds \cite{2013MNRAS.432..653D}. 

In this paper, we present an overview of the current methods and numerical corrections -- a `state of the union' between SPH and MHD -- and identify and consider some methods to push beyond the restrictions imposed by this. \cref{sec:background} provides an overview of how MHD and SPH are conventionally combined, and then in \cref{sec:momentum} and \cref{sec:induction} we discuss how the tensile instability is suppressed and several ways of preventing the growth of a monopole term.

\section{Magnetohydrodynamics in SPH}
\label{sec:background}

Cauchy's momentum equation is given by
\begin{equation}
\rmn{D}_t v^i = \frac{1}{\rho}\nabla^j\sigma^{ij} + g^i \text{~,}
\label{eqn:cauchymomeqn}
\end{equation}
where $\rmn{D}_t = \partial_t + v^i\nabla^i$ is the convective derivative operator (and $\partial_t$ the partial derivative with respect to time), $v^i$ represents the velocity, $\rho$ is the fluid density, $g^i$ represents any additional forces (in astrophysics usually self-gravity), Einstein's convention for summing over repeated indicies is used, and the stress tensor is represented by $\sigma^{ij}$. In non-magnetic fluids, 
\begin{equation}
\sigma^{ij} = -P\delta^{ij} \text{~,}
\end{equation}
where $\delta^{ij}$ is the Kronecker delta and $P$ is the fluid pressure, and \cref{eqn:cauchymomeqn} reduces to the usual form. However, in a magnetised fluid the tensor can be expanded to include the magnetic field so that
\begin{equation}
\sigma^{ij} = - \left( P + \frac{1}{2\mu_0} B^2 \right)\delta^{ij} + \frac{1}{\mu_0}B^iB^j \text{~,}
\end{equation}
where $B^i$ represents the magnetic field and $\mu_0$ the permeability of free space. The isotropic $\frac{1}{2\mu_0}B^2$ term is often referred to as magnetic `pressure' and the $\frac{1}{\mu_0}B^iB^j$ term as magnetic `tension'. This can be discretised in SPH (assuming the self-consistent variable smoothing lengths of \cite{2004MNRAS.348..139P} are used) as 
\begin{equation}
\rmn{D}_t v^i_a = \sum^N_b m_b \left( \frac{\sigma^{ij}_a}{\Omega_a\rho^2_a}\nabla^j_aW_{ab}\left(h_a\right) + \frac{\sigma^{ij}_b}{\Omega_b\rho^2_b}\nabla^j_aW_{ab}\left(h_b\right) \right)\text{~,} \label{eqn:momentum}
\end{equation}
where $a$ and $b$ represent SPH particles, $W_{ab}\left(h_{\{a,b\}}\right)$ is the smoothing kernel with smoothing lengths $h_{\{a,b\}}$, and $\Omega_{\{a,b\}}$ are terms to take account of gradients in $h$. This SPH discretisation exhibits perfect conservation properties, but as will be seen in \cref{sec:momentum} contains an inherent instability.

In addition to modifying the momentum equation, an equation for the evolution of the magnetic field must be solved. The induction equation (derived from the Maxwell-Faraday and Maxwell-Amp\'ere laws and the continuity equation) can be written in `Lagrangian' form as \cite{2012JCoPh.231..759P}
\begin{equation}
\rmn{D}_t \frac{B^i}{\rho} = \left( \frac{B^j}{\rho}\nabla^j \right)v^i \text{~,}
\end{equation}
which in SPH is discretised as
\begin{equation}
\rmn{D}_t \frac{B_a}{\rho_a} = -\sum^N_b \frac{m_b}{\Omega_a\rho^2_a} \left( v^i_a - v^i_b \right) B^j_a\nabla^j_aW_{ab}\left(h_a\right)\text{~.} 
\label{eqn:induction}
\end{equation}
The choice of operator is restricted to the antisymmetric derivative because of the requirement that the induction equation be invariant to the addition of a constant; this has consequences for other operator choices as discussed in \cite{2012JCoPh.231.7214T}. Alternatively, an induction equation based on 
\begin{equation}
\rmn{D}_t B^i = \left(B^j\nabla^j \right) v^i - B^i\left(\nabla^jv^j \right)
\end{equation}
discretised as
\begin{multline}
\rmn{D}_t B^i_a = -\frac{1}{\Omega_a\rho_a}\sum^N_b m_b \left( \left(v^i_a -v^i_b \right)B^j_a\nabla^j_aW_{ab}\left(h_a\right) \right. \\  -  \left(v^j_a -v^j_b \right)B^i_a\left.\nabla^j_aW_{ab}\left(h_a\right) \right)
\label{eqn:bvol}
\end{multline}
can be used, and in practice (although see the Appendix to \cite{2016MNRAS.455...51H}) is found to be little different to using \cref{eqn:momentum}. However, it is inherently somewhat inferior insofar as it involves an additional set of floating point operations introducing, at the very least, extra round-off error. We briefly note here the coupling between \cref{eqn:momentum} (which is the rate of change of $v^i$ as a function of \textit{inter alia} $B^i$) and \cref{eqn:induction} (which is the opposite). Any error in calculating one fluid parameter will then cause the evolution of the other to become incorrect, consequently MHD related errors can become disastrous very quickly.

As well as these equations, some form of artifical resistivity \cite{2013MNRAS.436.2810T} -- a similar concept to the ubiquitous artifical viscosity -- is needed to capture \diff{discontinuities} in the magnetic field structure. This is usually done by adding a term in the form of a `real' MHD resistivity, 
\begin{equation}
\rmn{D}_t \left. \frac{B^i_a}{\rho_a}\right|_\rmn{diss} = \eta \nabla^2 B_a \text{~,}
\end{equation}
but controlled by a switch similar artificial SPH viscosities. Excessive artificial resistivity can provide an illusion of increased stability\diff{ by virtue of a weaker magnetic field}.

As more fully discussed in \cref{sec:induction}, \diff{neither} \cref{eqn:induction} nor \cref{eqn:bvol} maintain the solenoidal constraint -- that $\nabla^iB^i = 0$ everywhere -- and therefore some modification is needed. How non-solenoidal the field has become is usually measured by the (dimensionless) parameter
\begin{equation}
\frac{h_a|\nabla^i_aB^i_a|}{|B^i|} \text{~,}
\end{equation}
which scales the divergence according to both the resolution and the field strength -- otherwise a very weak field may incorrectly appear to be much less incorrect that is actually the case. This issue, of course, is not unique to SPH or even to Langrangian methods in general.

This method of ideal MHD (or quasi-ideal if an artifical resistivity is used) can be extended into the non-ideal regime by adding terms for Ohmic resistivity, ambipolar diffusion -- both of which are dissipative -- and the non-dissipative Hall effect \cite{2008MNRAS.385.2269P}. Although this significantly expands the physical regimes that can be explored, none of these terms render the corrections discussed below nugatory (although in principle adding additional dissipative terms could temporarily forstall the need for a tensile instability correction).

\section{The Momentum Equation}
\label{sec:momentum}

In \cref{sec:background} we discussed how Cauchy's momentum equation is modified to add magnetic pressure and tension terms. The approach taken there has the advantage of being both correct in the continuum limit (\textit{i.e.} when $N_\rmn{ngh}\rightarrow\infty$) and exactly conserving angular and linear momentum. However, as we discussed in \cite{LBMP2015} last year, and more fully considered in \cite{1985MNRAS.216..883P}, this apparently simple formalism is susceptible to an instability -- rather similar to that seen in solid objects by \cite{1995JCoPh.116..123S}. If we write the magnetic tension component of the momentum equation as
\begin{equation}
\nabla^{j}_{a}B^{i}_{a}B^{j}_{a} = \left(B^{j}_{a}\nabla^{j}_{a}\right)B^{i}_a + B^{i}_{a}\left(\nabla^{j}_{a}B^{j}_{a}\right) \text{~,}
\end{equation}
we see that the second term -- $B^{i}_{a}\left(\nabla^{j}_{a}B^{j}_{a}\right)$ -- should always be zero due to Maxwell's Second Law. In a continuum fluid this is trivially satisfied, but in SPH the accuracy to which we maintain $\nabla^i_aB^i_a = 0$ is related to the accuracy of our derivative operators and the ordering of the particles under the stencil used to interpolate the derivative. As shown in \cref{eqn:momentum}, we use a symmetric derivative operator which, \textit{inter alia}, is not invariant to the addition of a constant term which magnifies this effect -- consistent with the observations of \cite{CHAUSSONNET2015} last year who found that the \textit{antisymmetric} operator ($G_{-}$ in their notation) was most consistent when applied to a perturbed lattice (although their overall SPH formalism was somewhat different to ours, these observations are comparable).

\begin{figure}
\centering{}
\includegraphics{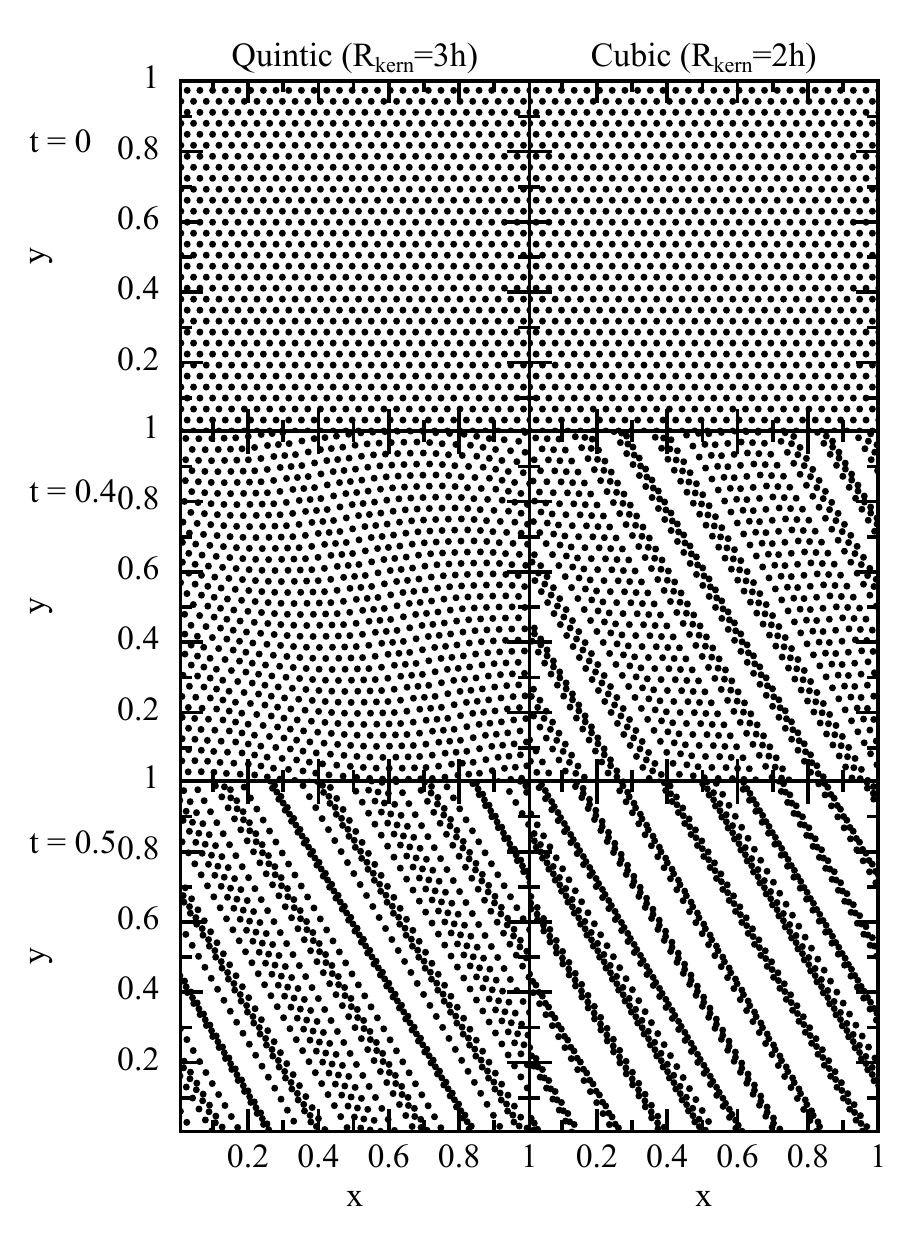}

\caption{Particle distribution\diff{, with no corrective term applied, }for a non-linear circularly polarized Alfv\'en wave propagating in 2.5D. This should be an exact solution to the equations of magnetohydrodynamics and as the wave moves the lattice should deform and then return to its original arrangement -- cf. \cref{fig:alfvenwavegood}. However, in SPMHD this calculation rapidly breaks down \diff{because it is entirely within the regime where }$\beta_\rmn{plasma} < 1$. Initially both kernels are consistent, but by $t = 0.4$ the `smaller' cubic B-spline has broken down. Finally at $t = 0.5$ the quintic kernel also fails.
\label{fig:alfvenwave}}
\end{figure}

Therefore, in all but a perfectly symmetrical lattice, we are actually solving an equation which looks like
\begin{equation}
\nabla^{j}_{a}B^{i}_{a}B^{j}_{a} = \left(B^{j}_{a}\nabla^{j}_{a}\right)B^{i}_a + B^{i}_{a}\left(\varrho_a\right)
\end{equation}
where $\varrho$ is the numerical monopole term caused by particle disorder. Even given this limitation, the stress tensor remains stable unless 
\begin{equation}
P_\rmn{mag} = \frac{1}{2\mu_0}B^2 > P_\rmn{hyd} \text{~,}
\end{equation}
\textit{i.e.} \diff{when} $\beta_\rmn{plasma} < 1$\diff{.} The monopole term acts along the magnetic field lines, where ordinarily the magnetic tension and pressure forces cancel out, and therefore causes particles to attract; however, because this spurious force is naturally proportional the field strength, any tensile pairing is suppressed when the fluid pressure is high enough\footnotemark{}. Consequently, for many calculations -- those where $\beta_\rmn{plasma} > 1$ is always true -- this effect can be ignored, although, as seen in \cref{fig:alfvenwave}, when this effect does take hold it rapidly destroys the calculation. Further, by switching between a cubic and quintic B-spline kernel \cite{1985JCoPh..60..253M}, the effect of different stencils can be seen: the smaller $2h$ compact support radius for the cubic spline becomes unstable more rapidly than the $3h$ quintic function. 

\footnotetext{This observation led \cite{2004MNRAS.348..123P} to propose a short-range repulsive force to prevent pairing when $\beta_\rmn{plasma} < 1$, but it is difficult to determine what is `short-range' when variable smoothing lengths are used.}

\begin{figure}
\centering{}
\includegraphics{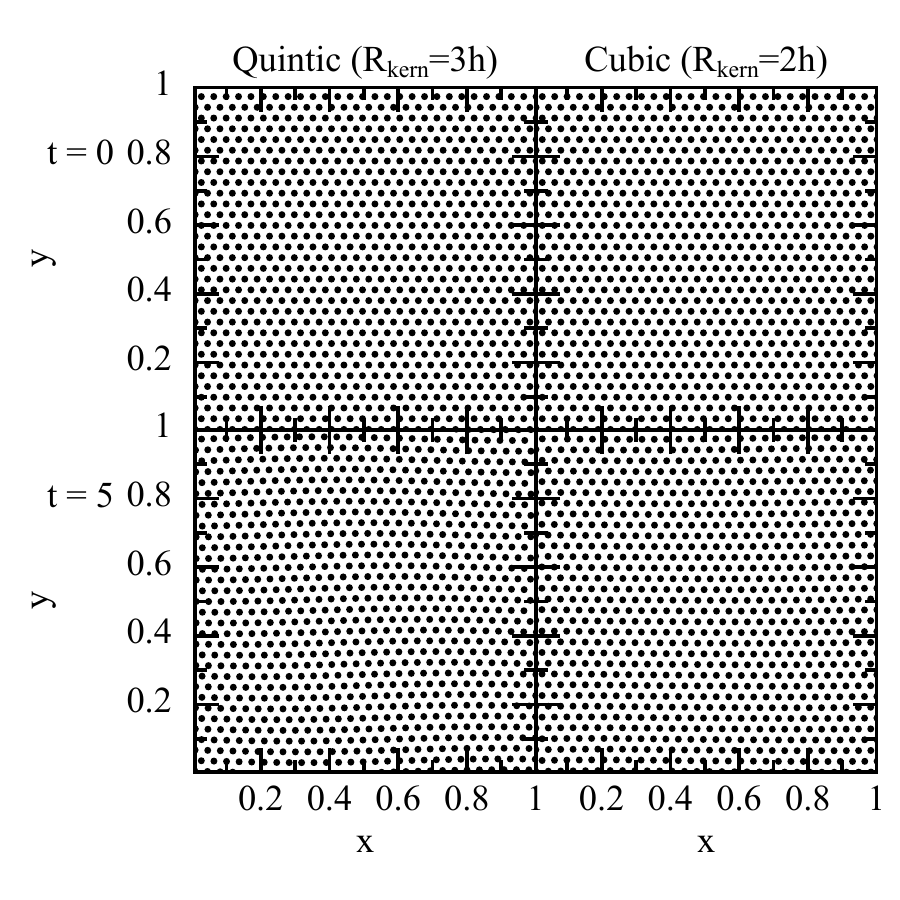}

\caption{A non-linear circularly polarized Alfv\'en wave in 2.5D -- the same setup as in \cref{fig:alfvenwave}. For either choice of kernel, with the source term correction from \cite{2001ApJ...561...82B}, this calculation is stable.
\label{fig:alfvenwavegood}}
\end{figure}

Very \diff{low} $\beta_\rmn{plasma}$ \diff{values} are common in astrophysical calculations -- for example in \cite{2015MNRAS.451.4807L} values of $< 10^{-2}$ were obtained. Some method of preventing particles from unphysically attracting each other is therefore essential in all but the most trivial of calculations. For many years, the method proposed by \cite{2001ApJ...561...82B} whereby a source term `equal' to the error is subtracted was used successfully. In this method, we in effect solve
\begin{equation}
\nabla^{j}_{a}B^{i}_{a}B^{j}_{a} = \left(B^{j}_{a}\nabla^{j}_{a}\right)B^{i}_a + B^{i}_{a}\left(\varrho_a\right) -  \chi{}B^{i}_{a}\left(\widetilde{\varrho_a}\right) \text{~,}
\end{equation}
where $\widetilde{\varrho_a} = \nabla^i_aB^i_a$ is our estimate (calculated using the SPH divergence operator) of the numerical monopole term and $\chi \in [0,~1]$ is the degree of correction applied. This corrective term naturally makes the SPMHD equations no longer conserve linear momentum, so a value of $\chi = \frac{1}{2}$ was originally proposed \cite{2004ApJS..153..447B} as a minimum value to obtain stability; subsequent experience \cite{2012JCoPh.231.7214T} indicated that $\chi = 1$ was necessary for all but the simplest test problems. Clearly, the utility of this method is limited to how well $\widetilde{\varrho_a}$ can be calculated -- and we must use a symmetric operator for consistency so this is an inherently noisy operation\footnotemark{}.

\footnotetext{In the continuum limit, $\varrho_a \equiv \widetilde{\varrho_a} = 0$, so this correction would be safe but unnecessary.}

\begin{figure*}
\centering{}
\includegraphics{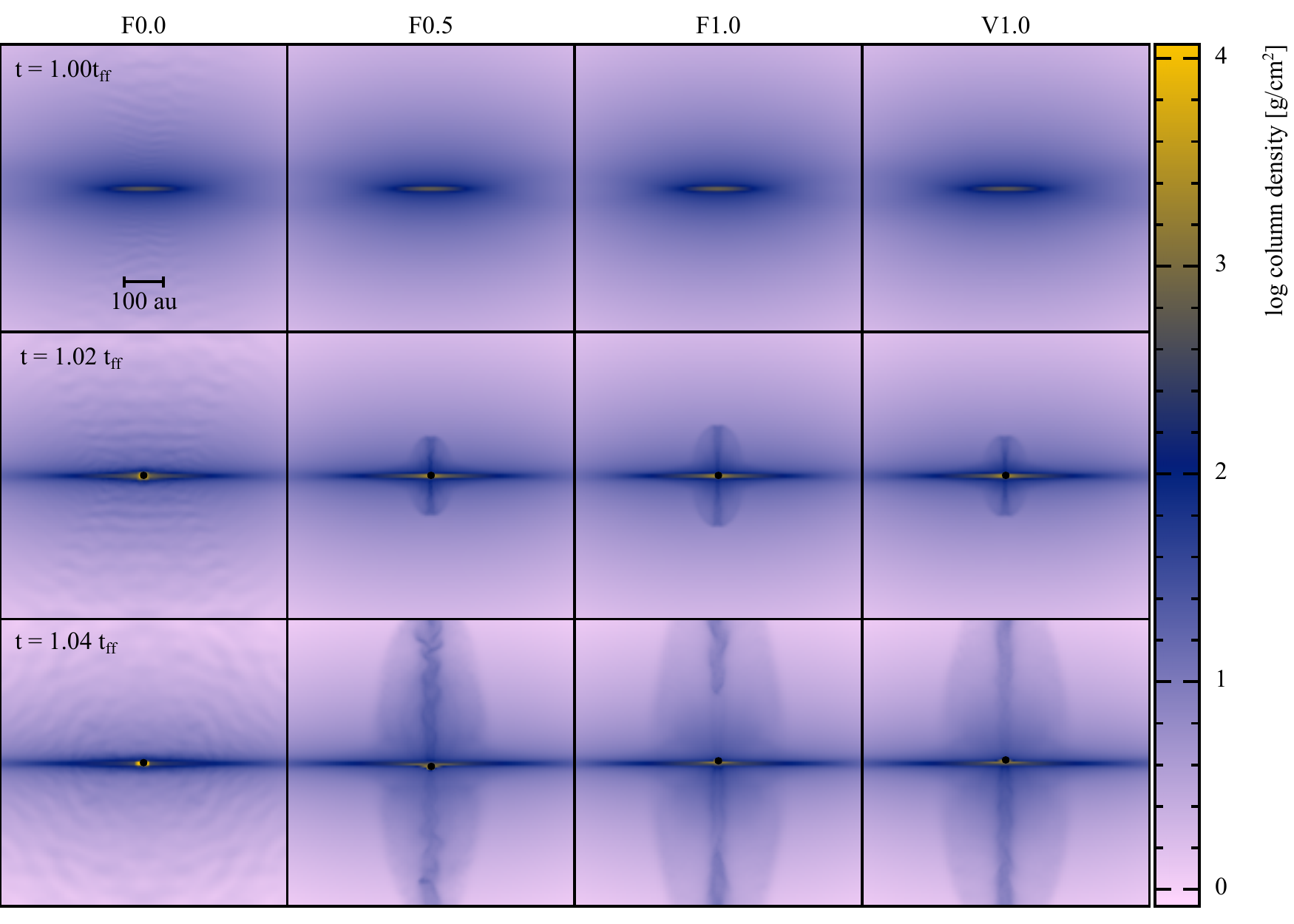}

\caption{Four protostellar collapse simulations (the initial conditions being the same as in both \cite{LBMP2015} and \cite{2015MNRAS.451.4807L}, \textit{viz.} a $\approx 10~\text{K}$ $1~\rmn{M}_\odot$ sphere \diff{in solid body rotation} initially in pressure equilibrium with a warm low density, \diff{non-rotating,} container medium, with an initial magnetic field aligned with the $z-$axis with a mass-to-flux ratio \cite{1976ApJ...210..326M} of $\mu = 5$). F-numbers denote fixed values of $\chi$, so that F0.0 represents $\chi = 0$ and so on, and the V1.0 calculation denotes a switch like that in \cref{eqn:chiswitch} with $\chi_\rmn{max} = 1$, the similar $\chi_\rmn{max} = \frac{1}{2}$ is not shown. The outflow velocity and morphology clearly correlates with $\chi$, although the particle distribution for $\chi = 0$ is also incorrect due to tensile pairing, evidenced by the `wave-like' structures seen above and below the pseudo-disc. The sink particle which represents the protostar for $\chi = \frac{1}{2}$ and $1$ has begun to fall out of the pseudo-disc \diff{due to the failure of momentum conservation}.
\label{fig:protostar}}
\end{figure*}

\begin{figure}
\centering{}
\includegraphics{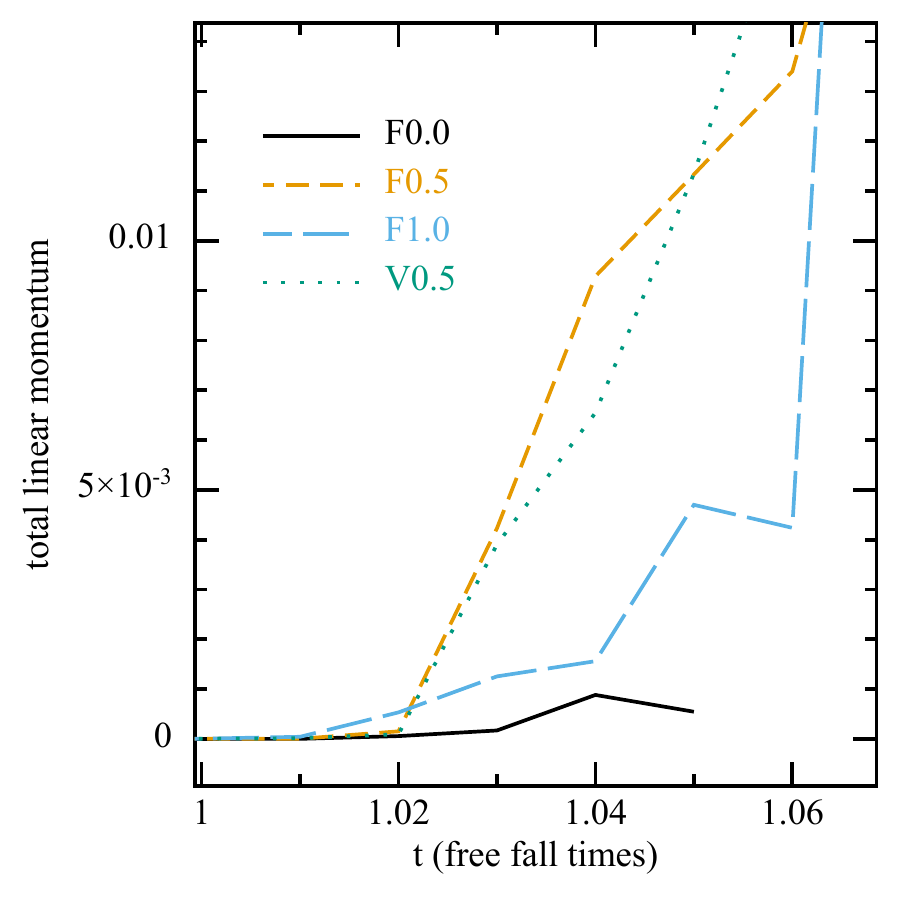}

\caption{Linear momentum against time for the calculations presented in \cref{fig:protostar}. All values of $\chi > 0$ including the variable $\chi$ calculation show non-trivial increase in linear momentum correlated to the moment when the protostar has separated from the pseudo-disc; $\chi = 0$ maintains the exact SPH momentum conservation at the expense of an unphysical particle distribution.
\label{fig:protostarlinmom}}
\end{figure}

As seen in \cref{fig:alfvenwavegood}, this correction can be remarkably effective (the calculation presented there can essentially be run \textit{ad infinitum} without any deleterious effects) provided the particle disorder is sufficiently small -- and hence $\varrho_a \sim \widetilde{\varrho_a}$. We have found that the degree of particle disorder experienced in a star formation calculation can rapidly cause significant errors, which limited the time after sink particle \cite{1997MNRAS.288.1060B} insertion these calculations could be run -- the calculation being correct, like the circularly polarized Alfv\'en wave, until the $\widetilde{\varrho_a}$ related error manifestly exceeds the inherent $\varrho_a$ error.
One way of \diff{limiting} these conservation related errors would be to use a switch on $\chi$, an approach which was first proposed by \cite{2004ApJS..153..447B} and was used by \cite{2016MNRAS.457.1037W} \diff{with some success}. Several switches have been proposed, here we use a simple particle based switch given by 
\begin{equation}
\chi_a = \rmn{MIN}\left(\chi_\rmn{max}\left(1 - \beta_\rmn{plasma} \right), 0\right) \text{~,}
\label{eqn:chiswitch}
\end{equation}
where $\chi_\rmn{max} \in [0, 1]$ like the earlier fixed values. This allows the correction to entirely vanish when $\beta_\rmn{plasma} > 1$ (where the equations are stable against the instability anyway) and to rapidly switch on as $\beta_\rmn{plasma}$ falls below unity. Calculations with $\chi_\rmn{max} = \frac{1}{2}$ and 1 were performed and no substantial improvement was obtained over simply using a fixed value. 
In \cref{fig:protostar} we show the effect of these errors on \diff{a star} formation problem -- the failure of momentum conservation causes the protostar to fall out of the surrounding pseudo-disc -- and in \cref{fig:protostarlinmom} the correlation of this unphysical effect with the failure to conserve momentum is shown. 

We also show the effect of setting $\chi = 0$ where the pseudo-disc remains intact. However, the particle distribution exhibits clear errors which can be seen in the column density as `waves' caused by particles clumping as the sphere collapses, so running with no corrective term is also not an option. 
\diff{Outflows are obtained as long as} $\chi > 0$\diff{, but the exact speeds and morphologies vary with differing values of }$\chi$\diff{.}
Tensile pairing becomes more problematic in areas of higher field strengths and increased particle disorder, exactly the same conditions present when a bipolar jet is being formed from a wound-up magnetic field. Therefore, the likely cause of the outflow suppression is that particles which would otherwise form the jet are being stuck together similar to the lines seen in the test problem.

Naively, we would assume that the error introduced by particle disorder on the $\nabla^j_aB^i_aB^j_a$ operator is the same as that on the $\nabla^i_aB^i$ operator. However, these results indicate that this is not the case -- indeed, the divergence operator may be overestimating the error when the particle disorder is extreme coupled with a strong magnetic field, this presents a severe limitation on how long calculations can be run after the field has become sufficiently complicated. An overestimate of the error can be imagined as actually super-imposing a monopole term onto the momentum equation but with the sign reversed, a clearly undesirable prospect. In effect we are left with a trade off between avoiding an instability due to a non-vanishing $\nabla^iB^i$ term and errors due to non-exact conservation of \diff{momentum.}

\section{The Induction Equation}
\label{sec:induction}

The direct induction equation presented in \cref{sec:intro} is not the only way to calculate the evolution of the magnetic field (although it is the simplest). A major flaw is that it does not preseve the solenoidal constraint -- although for different reasons to those outlined in \cref{sec:momentum}. Being an antisymmetric operator, \cref{eqn:induction} is much less susceptible to numerical noise due to particle disorder \cite{CHAUSSONNET2015}, instead floating-point round-off error takes over. As formulated, the constraint 
\begin{equation}
\nabla^i B^i \equiv 0
\label{eqn:divbconstraint}
\end{equation}
does not enter the equations of SPMHD -- although it can be trivially imposed as an initial condition -- and the related constraint
\begin{equation}
\rmn{D}_t \nabla^i B^i = 0
\label{eqn:roundoff}
\end{equation}
is not in general true for any discretised equation. Consequently the field will naturally become non-solenoidal as numerical errors add up\footnotemark{}. Unlike the issues discussed in \cref{sec:momentum}, keeping the induction equation well-behaved is not unique to SPH. In recent years this has been solved by adopting a hyperbolid divergence cleaning approach from grid codes \cite{2002JCoPh.175..645D}. In this, an additional scalar field $\psi$, evolved according to \cite{2012JCoPh.231.7214T}
\begin{equation}
\rmn{D}_t\phi = -c^2_\rmn{c}\nabla^iB^i - \frac{\phi}{\tau} - \frac{\phi \nabla^iv^i}{2} \text{~,}
\label{eqn:divbwave}
\end{equation}
where the cleaning wave speed, $c_\rmn{c}$, is invariably set to the magnetosonic wave speed and the damping timescale,
\begin{equation}
\tau = \frac{h}{c_\rmn{c}\sigma} \text{~,}
\end{equation}
set to dissipate the wave over a small number of smoothing lengths, for instance $\sigma = 0.8$ in \cite{LBMP2015}.

This damped wave equation is then coupled to the induction equation -- like the artificial resistivity -- by adding a term of the form 
\begin{equation}
\rmn{D}_t \left. \frac{B^i}{\rho} \right|_\rmn{cleaning} = -\nabla^i\psi \text{~.}
\end{equation}
dissipating any unphysical monopole growth. This successfully suppresses the growth of any monopoles, although we have observed issues where the information speed in the fluid is faster than the cleaning wave speed -- in effect the errors can escape from the cleaning -- as happens in rapidly rotating accretion discs. There is an inherent risk in this approach that the solenoidal field produced by the cleaning wave will not be the \textit{correct} field -- that is that the cleaning has itself caused a change in the field geometry.

\footnotetext{Obviously test problems could be contrived where the floating-point errors cancel, but any realistic calculation will reasonably quickly generate a non-trivial $\nabla^i B^i$.}

Solving for the rate-of-change of $B^i$ is not the only way to evolve a magnetic field: alternatively, the magnetic vector potential,
\begin{equation}
B^i = \epsilon^{ijk}\nabla^jA^k \text{~,}
\label{eqn:vecpot}
\end{equation}
where $\epsilon^{ijk}$ represents the totally antisymmetric Levi-Civita tensor, can be evolved. This has the advantage that \cref{eqn:divbconstraint} becomes a constraint in the SPMHD formalism since $\nabla^i \epsilon^{ijk} \nabla^jA^k = 0$ by definition, neatly sidestepping the round-off errors in \cref{eqn:roundoff}. However, an additional complication is that there are infinitely many vector potentials than can produce the same magnetic field, since 
\begin{equation}
\epsilon^{ijk} \nabla^j \nabla^k \phi = 0 \text{~,}
\end{equation}
$\phi$ being any continuously differentiable scalar field, one can always write
\begin{equation}
A^{k'} = A^{k} + \nabla^k \phi
\end{equation}
and obtain the same resultant magnetic field. Choice of gauge -- and ensuring that additional error is not introduced either by this choice or by the SPH operators used to maintain -- thereby becomes important. In \cite{2010MNRAS.401.1475P}, a Galilean invariant gauge whereby 
\begin{equation}
\phi = v^iA^i \text{~,}
\label{eqn:gaugeaxel}
\end{equation}
ensures perfect momentum conservation. Alternative gauges include the Weyl gauge, 
\begin{equation}
\phi = 0 \text{~,}
\end{equation}
and the Modified Psuedo-Lorenz gauge\footnotemark{} (MPLG) proposed by \cite{2015JCoPh.282..148S}, where
\begin{equation}
\partial_t \phi = -c^2\nabla^iA^i \text{~,}
\label{eqn:gaugemplg}
\end{equation}
similar to \cref{eqn:divbwave} $c$ is set to the fastest magnetosonic wave speed.
The vector potential would then be evolved by either 
\begin{equation}
\rmn{D}_t A^i = v^j\nabla^iA^j + \nabla^i\phi \text{~or}
\end{equation}
\begin{equation}
\rmn{D}_t A^i = -A^j\nabla^i v^j + \nabla^i\phi
\label{eqn:aindgal}
\end{equation}
for the Galilean invariant gauge (the SPH operators for which are given in \cite{2010MNRAS.401.1475P}, and where the contributions from external fields and dissipative terms have been neglected).

\footnotetext{The psuedo-Lorenz gauge is, in effect, a Lorenz gauge where the wave speed is some arbitrarily chosen velocity (\textit{e.g.} the speed of sound) which is less than the speed of light.}

\begin{figure}
\centering{}
\begin{minipage}{.50\textwidth}
\centering{}
\begin{center}
\includegraphics[width=.45\textwidth]{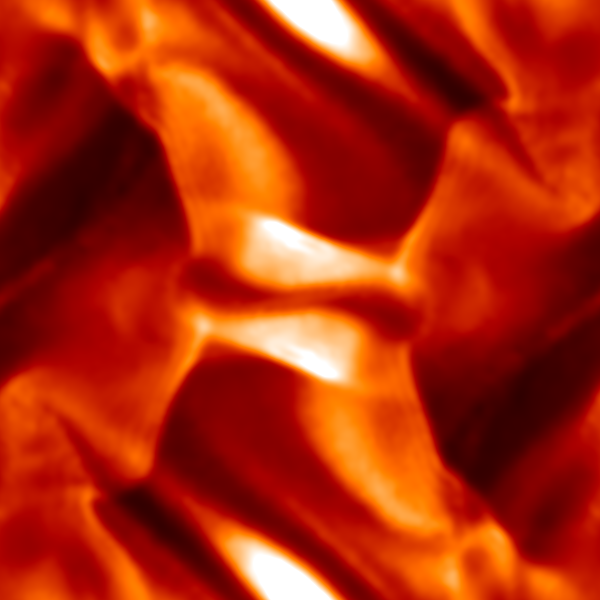}\includegraphics[width=.45\textwidth]{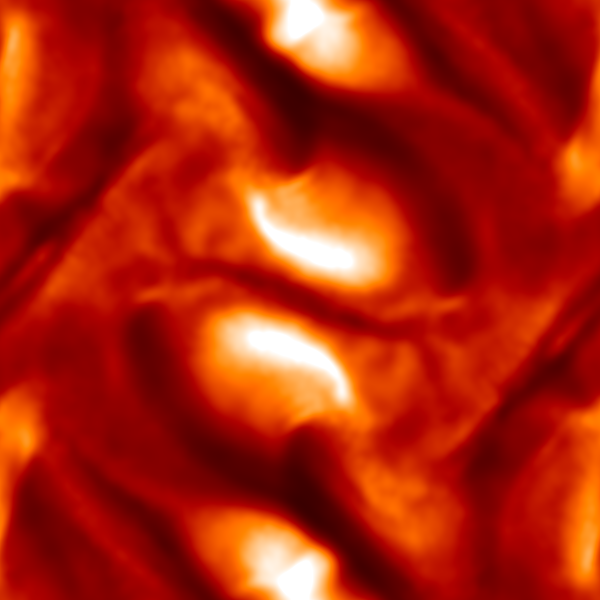}
\includegraphics[width=.45\textwidth]{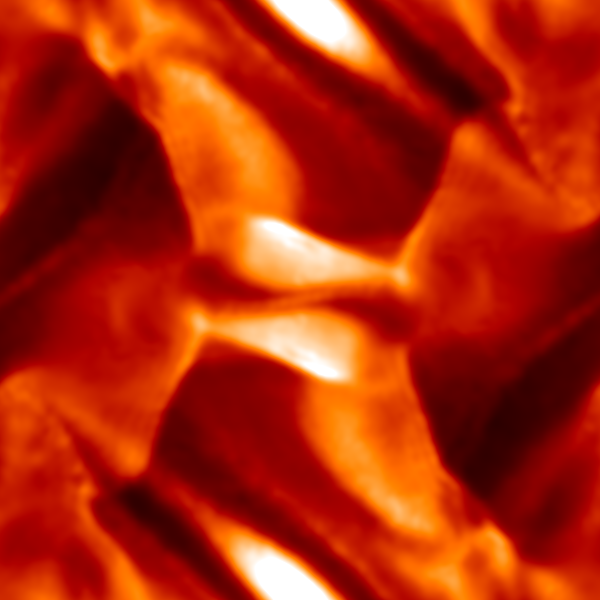}\includegraphics[width=.45\textwidth]{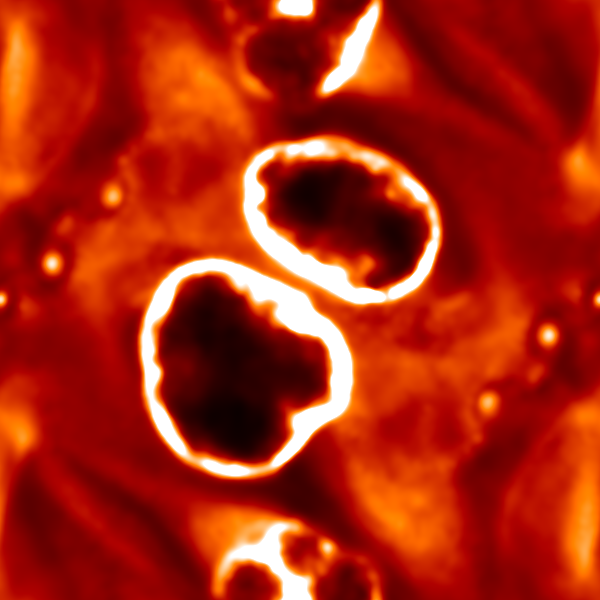}

\caption{Orszag \& Tang \cite{1979JFM....90..129O} test vorticies at $t = 0.5$ and $t = 0.79$ (left and right columns) for direct induction using \cref{eqn:induction} and the vector potential with a Gallilean invarient gauge \cref{eqn:aindgal}. No corrective term (see \cref{sec:momentum}) is applied to the direct induction momentum equation yet the vortex remains stable; however, the vector potential has become significantly disrupted and formed large voids characteristic of tensile pairing.
\label{fig:otvec}}
\end{center}
\end{minipage}
\end{figure}

\begin{figure}
\centering{}
\includegraphics[width=\columnwidth]{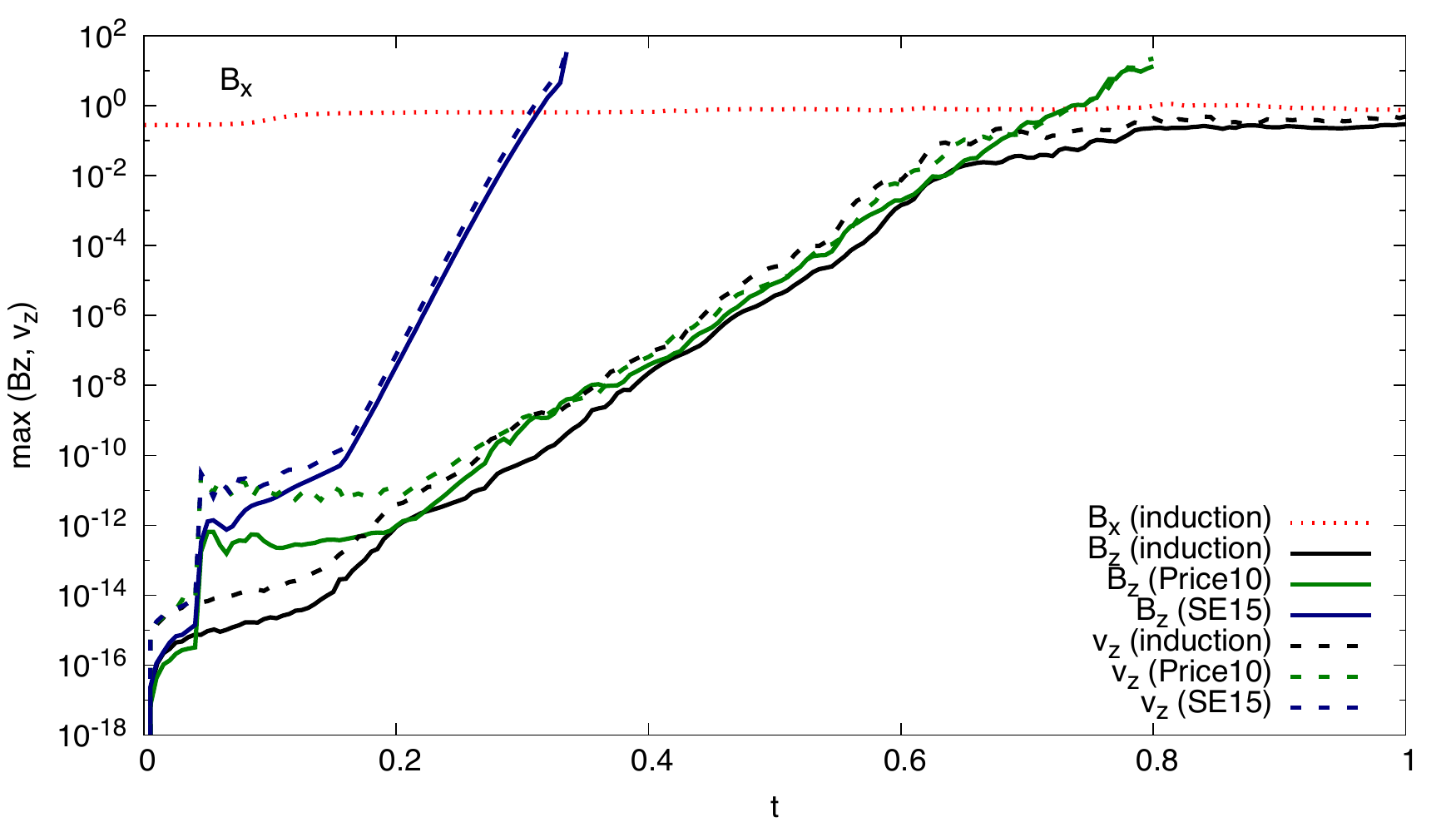}

\caption{The $z-$components of the magnetic field ($B_z$) and velocity field ($v_z$) in an Orszag-Tang \cite{2001ApJ...561...82B} test vortex for the normal direct induction (black lines) equation (\cref{eqn:induction}), the Galilean invariant (Price10, green lines) gauge (\cref{eqn:gaugeaxel}) and the MPL (SE15, blue lines) gauge (\cref{eqn:gaugemplg}). We include the direct induction $B_x$ component for reference. As particles deviate from the lattice in the $z-$direction simply due to numerical noise, a spurious $v_z$ component is produced which then causes the growth of $B_z$ -- and hence further $v_z$ --  in all three formalisms. Although the Price10 and direct induction methods agree until about $t = 0.7$, the vector potential form does not level off as $B_z \rightarrow B_x$. The SE10 method is significantly less stable and the $v_z$ and $B_z$ terms grown much more rapidly.
\label{fig:terry}}
\end{figure}

The gauge constraint in \cref{eqn:gaugeaxel} coupled with \cref{eqn:aindgal} would appear to be the natural choice for SPH, and indeed, does provide perfect conservation properties and puts the SPH derivative operator on the velocity in a similar way to \cref{eqn:induction}. However, like all other gauge choices it is highly unstable: particle disorder rapidly causes the potential to increase nonphysically -- and hence the magnetic field shown in \cref{fig:terry}. In addition, a self-consistent tensor force derived from this gauge choice is \textit{even more} susceptible to the tensile instability -- as shown in \cref{fig:otvec} -- and therefore requires more corrective action, and consequently inferior conservation properties to just using direct induction.

A potential alternative approach was provided by \cite{2015JCoPh.282..148S}, which involved the combination of the gauge given by \cref{eqn:gaugemplg} with a form of `divergence cleaning' to maintain the constraint that $\nabla^i A^i = 0$, which is aided by the fact the Lorenz gauge condition is in effect a wave equation, and a magnetic field smoothing term. Whilst promising, we find that this still suffers from many of the pitfalls of previous gauges and is only stable in the O-T vortex when the combination of the cleaning and smoothing keep the $A_{x,y}$ components less than $A_{z}$ -- otherwise, as shown in \cref{fig:terry}, it rapidly becomes unphysical. 

In addition to any errors due to failure to maintain the gauge constraint, there is an universal issue related to the second step of any formalism which evolves $A^i$, namely that the magnetic $B^i$ field must then be determined by a second operation -- in principle some form of direct second derivative could be used, but second derivatives in SPH are notoriously difficult to implement correctly. Given that particle disorder is already known to be deleterious even for the induction equation, performing a second SPH operation causes additional error. In effect the error from calculating $\rmn{D}_t A^i$ is compounded by the error from turning that into a $B^i$-field via \cref{eqn:vecpot} in order to calculate the force.

\section{Conclusion}
\label{sec:conclusion}

Notwithstanding the inherent limitations, smoothed particle magnetohydrodynamics can be applied \diff{to} a significant range of problems. Importantly, calculations \diff{where }$\beta_\rmn{plasma} > 1$\diff{ throughout }are markedly simpler than the converse. Even then, if the particle lattice remains reasonably well ordered the source term correction is robust at removing a numerical monopole term and preventing a tensile instability. However, in large scale calculations with complicated -- often turbulent -- particle distributions the momentum error introduced by this correction can ultimately limit the duration of the calculation.

Provided some form of divergence cleaning is used, the induction equation presents a smaller range of limitations. A vector potential formalism would be preferable as this implicitly enforces Maxwell's second law. Particle disorder, however, has so far stymied any attempt at a robust vector potential method because of the need to perform an additional set of (noise sensitive) SPH derivative operations. 


\section*{Acknowledgment}
Test calculations (shown in \cref{fig:alfvenwave,fig:alfvenwavegood,fig:otvec}) were performed using D.J. Price's \texttt{ndspmhd} code; protostellar collapse calculations (show in \cref{fig:protostar}) were performed using the hybrid \texttt{MPI} and \texttt{OpenMP} code \texttt{sphNG}, originally written by W. Benz \cite{X} modified and significantly expanded by M.R. Bate and collaborfators \cite{1997MNRAS.288.1060B,2014MNRAS.437...77B}. Rendered plots were then produced using the \texttt{splash} \cite{2007PASA...24..159P} visualisation programme. 

This work used the DiRAC Complexity system, operated by the University of Leicester IT Services, which forms part of the STFC DiRAC HPC Facility (www.dirac.ac.uk). This equipment is funded by BIS National E-Infrastructure capital grant ST/K000373/1 and STFC DiRAC Operations grant ST/K0003259/1. DiRAC is part of the National E-Infrastructure.

Calculations were also performed on the University of Exeter Supercomputer,a DiRAC facility jointly funded by the STFC, the Large Facilities Capital Fund of BIS and the University of Exeter.

BTL acknowledges support from an STFC Studentship. This work was also supported by the European Research Council under the European Community's Seventh Framework Programme (\diff{FP7/2007-2013} Grant Agreement No. 339248). TST is supported by a CITA Post-doctoral Research Fellowship.



%

\bibliographystyle{IEEEtran.bst}
\bibliography{IEEEabrv,MHD_SotU}

\begin{thebibliography}{10}
\providecommand{\url}[1]{#1}
\csname url@samestyle\endcsname
\providecommand{\newblock}{\relax}
\providecommand{\bibinfo}[2]{#2}
\providecommand{\BIBentrySTDinterwordspacing}{\spaceskip=0pt\relax}
\providecommand{\BIBentryALTinterwordstretchfactor}{4}
\providecommand{\BIBentryALTinterwordspacing}{\spaceskip=\fontdimen2\font plus
\BIBentryALTinterwordstretchfactor\fontdimen3\font minus
  \fontdimen4\font\relax}
\providecommand{\BIBforeignlanguage}[2]{{%
\expandafter\ifx\csname l@#1\endcsname\relax
\typeout{** WARNING: IEEEtran.bst: No hyphenation pattern has been}%
\typeout{** loaded for the language `#1'. Using the pattern for}%
\typeout{** the default language instead.}%
\else
\language=\csname l@#1\endcsname
\fi
#2}}
\providecommand{\BIBdecl}{\relax}
\BIBdecl

\bibitem{1977AJ.....82.1013L}
L.~B. {Lucy}, ``{A numerical approach to the testing of the fission
  hypothesis},'' \emph{\aj}, vol.~82, pp. 1013--1024, Dec. 1977.

\bibitem{1977MNRAS.181..375G}
R.~A. {Gingold} and J.~J. {Monaghan}, ``{Smoothed particle hydrodynamics -
  Theory and application to non-spherical stars},'' \emph{\mnras}, vol. 181,
  pp. 375--389, Nov. 1977.

\bibitem{1949Sci...109..165H}
W.~A. {Hiltner}, ``{Polarization of Light from Distant Stars by Interstellar
  Medium},'' \emph{Science}, vol. 109, p. 165, Feb. 1949.

\bibitem{1985MNRAS.216..883P}
G.~J. {Phillips} and J.~J. {Monaghan}, ``{A numerical method for
  three-dimensional simulations of collapsing, isothermal, magnetic gas
  clouds},'' \emph{\mnras}, vol. 216, pp. 883--895, Oct. 1985.

\bibitem{1995JCoPh.116..123S}
J.~W. {Swegle}, D.~L. {Hicks}, and S.~W. {Attaway}, ``{Smoothed Particle
  Hydrodynamics Stability Analysis},'' \emph{Journal of Computational Physics},
  vol. 116, pp. 123--134, Jan. 1995.

\bibitem{2009MNRAS.398...33P}
D.~J. {Price} and M.~R. {Bate}, ``{Inefficient star formation: the combined
  effects of magnetic fields and radiative feedback},'' \emph{\mnras}, vol.
  398, pp. 33--46, Sep. 2009.

\bibitem{2014MNRAS.437...77B}
M.~R. {Bate}, T.~S. {Tricco}, and D.~J. {Price}, ``{Collapse of a molecular
  cloud core to stellar densities: stellar-core and outflow formation in
  radiation magnetohydrodynamic simulations},'' \emph{\mnras}, vol. 437, pp.
  77--95, Jan. 2014.

\bibitem{2015MNRAS.451.4807L}
B.~T. {Lewis}, M.~R. {Bate}, and D.~J. {Price}, ``{Smoothed particle
  magnetohydrodynamic simulations of protostellar outflows with misaligned
  magnetic field and rotation axes},'' \emph{\mnras}, vol. 451, pp. 4807--4818,
  Jul. 2015.

\bibitem{LewisSubmit}
------, ``{The dependence of protostar formation on the geometry and strength
  of the initial magnetic field},'' 2016, submitted.

\bibitem{2007MNRAS.379..915R}
S.~{Rosswog} and D.~{Price}, ``{MAGMA: a three-dimensional, Lagrangian
  magnetohydrodynamics code for merger applications},'' \emph{\mnras}, vol.
  379, pp. 915--931, Aug. 2007.

\bibitem{2013MNRAS.432..653D}
C.~L. {Dobbs} and J.~E. {Pringle}, ``{The exciting lives of giant molecular
  clouds},'' \emph{\mnras}, vol. 432, pp. 653--667, Jun. 2013.

\bibitem{2004MNRAS.348..139P}
D.~J. {Price} and J.~J. {Monaghan}, ``{Smoothed Particle Magnetohydrodynamics -
  II. Variational principles and variable smoothing-length terms},''
  \emph{\mnras}, vol. 348, pp. 139--152, Feb. 2004.

\bibitem{2012JCoPh.231..759P}
D.~J. {Price}, ``{Smoothed particle hydrodynamics and magnetohydrodynamics},''
  \emph{Journal of Computational Physics}, vol. 231, pp. 759--794, Feb. 2012.

\bibitem{2012JCoPh.231.7214T}
T.~S. {Tricco} and D.~J. {Price}, ``{Constrained hyperbolic divergence cleaning
  for smoothed particle magnetohydrodynamics},'' \emph{Journal of Computational
  Physics}, vol. 231, pp. 7214--7236, Aug. 2012.

\bibitem{2016MNRAS.455...51H}
P.~F. {Hopkins} and M.~J. {Raives}, ``{Accurate, meshless methods for
  magnetohydrodynamics},'' \emph{\mnras}, vol. 455, pp. 51--88, Jan. 2016.

\bibitem{2013MNRAS.436.2810T}
T.~S. {Tricco} and D.~J. {Price}, ``{A switch to reduce resistivity in smoothed
  particle magnetohydrodynamics},'' \emph{\mnras}, vol. 436, pp. 2810--2817,
  Dec. 2013.

\bibitem{2008MNRAS.385.2269P}
B.~P. {Pandey} and M.~{Wardle}, ``{Hall magnetohydrodynamics of partially
  ionized plasmas},'' \emph{\mnras}, vol. 385, pp. 2269--2278, Apr. 2008.

\bibitem{LBMP2015}
B.~T. {Lewis}, M.~R. {Bate}, J.~J. {Monaghan}, and D.~P. {Price}, ``{Stable
  smoothed particle magnetohydrodynamics in very steep density gradients},'' in
  \emph{10th international SPHERIC workshop}, Parma, Italy, Jun. 2015.

\bibitem{CHAUSSONNET2015}
G.~{Chaussonnet}, S.~{Braun}, L.~{Wieth}, R.~{Koch}, and H.-J. {Bauer},
  ``{Influence of particle disorder and smoothing length on SPH operator
  accuracy},'' in \emph{10th international SPHERIC workshop}, Parma, Italy,
  Jun. 2015.

\bibitem{1985JCoPh..60..253M}
J.~J. {Monaghan}, ``{Extrapolating B. Splines for Interpolation},''
  \emph{Journal of Computational Physics}, vol.~60, pp. 253--262, Sep. 1985.

\bibitem{2004MNRAS.348..123P}
D.~J. {Price} and J.~J. {Monaghan}, ``{Smoothed Particle Magnetohydrodynamics -
  I. Algorithm and tests in one dimension},'' \emph{\mnras}, vol. 348, pp.
  123--138, Feb. 2004.

\bibitem{2001ApJ...561...82B}
S.~{B{\o}rve}, M.~{Omang}, and J.~{Trulsen}, ``{Regularized Smoothed Particle
  Hydrodynamics: A New Approach to Simulating Magnetohydrodynamic Shocks},''
  \emph{\apj}, vol. 561, pp. 82--93, Nov. 2001.

\bibitem{2004ApJS..153..447B}
------, ``{Two-dimensional MHD Smoothed Particle Hydrodynamics Stability
  Analysis},'' \emph{\apjs}, vol. 153, pp. 447--462, Aug. 2004.

\bibitem{1997MNRAS.288.1060B}
M.~R. {Bate} and A.~{Burkert}, ``{Resolution requirements for smoothed particle
  hydrodynamics calculations with self-gravity},'' \emph{\mnras}, vol. 288, pp.
  1060--1072, Jul. 1997.

\bibitem{2016MNRAS.457.1037W}
J.~{Wurster}, D.~J. {Price}, and M.~R. {Bate}, ``{Can non-ideal
  magnetohydrodynamics solve the magnetic braking catastrophe?}''
  \emph{\mnras}, vol. 457, pp. 1037--1061, Mar. 2016.

\bibitem{1976ApJ...210..326M}
T.~C. {Mouschovias} and L.~{Spitzer}, Jr., ``{Note on the collapse of magnetic
  interstellar clouds},'' \emph{\apj}, vol. 210, p. 326, Dec. 1976.

\bibitem{2002JCoPh.175..645D}
A.~{Dedner}, F.~{Kemm}, D.~{Kr{\"o}ner}, C.-D. {Munz}, T.~{Schnitzer}, and
  M.~{Wesenberg}, ``{Hyperbolic Divergence Cleaning for the MHD Equations},''
  \emph{Journal of Computational Physics}, vol. 175, pp. 645--673, Jan. 2002.

\bibitem{2010MNRAS.401.1475P}
D.~J. {Price}, ``{Smoothed Particle Magnetohydrodynamics - IV. Using the vector
  potential},'' \emph{\mnras}, vol. 401, pp. 1475--1499, Jan. 2010.

\bibitem{2015JCoPh.282..148S}
F.~A. {Stasyszyn} and D.~{Elstner}, ``{A vector potential implementation for
  smoothed particle magnetohydrodynamics},'' \emph{Journal of Computational
  Physics}, vol. 282, pp. 148--156, Feb. 2015.

\bibitem{1979JFM....90..129O}
S.~A. {Orszag} and C.-M. {Tang}, ``{Small-scale structure of two-dimensional
  magnetohydrodynamic turbulence},'' \emph{Journal of Fluid Mechanics},
  vol.~90, pp. 129--143, Jan. 1979.

\bibitem{X}
W.~{Benz}, in \emph{Numerical Modelling of Nonlinear Stellar Pulsations
  Problems and Prospects}, J.~R. {Buchler}, Ed.\hskip 1em plus 0.5em minus
  0.4em\relax Dordrecht: Kluwer, 1990, p. p. 269.

\bibitem{2007PASA...24..159P}
D.~J. {Price}, ``{splash: An Interactive Visualisation Tool for Smoothed
  Particle Hydrodynamics Simulations},'' \emph{\pasa}, vol.~24, pp. 159--173,
  Oct. 2007.

\end{thebibliography}

\end{document}